\documentclass{aa} 

\usepackage{fleqn}

\usepackage{url}
 
\usepackage{graphicx}
 
\usepackage{natbib}
\bibpunct{(}{)}{;}{a}{}{,}
\newcommand\oao{OAO~1657$-$415}

\begin{document}

\title{INTEGRAL observations of \oao: gamma-ray tomography   of
a B supergiant}
\titlerunning{INTEGRAL observations of \oao}
\authorrunning{M.~Denis, T.~Bulik, R.~Marcinkowski}
\author{M.~Denis \inst{1},
        T.~Bulik \inst{2},
        R.~Marcinkowski \inst{1}
       }

\institute{
 $^1$ Space Research Center, Bartycka 18A, 00-716 Warsaw, Poland\\
 $^2$ Nicolaus Copernicus Astronomical Center, Bartycka 18, 00-716 Warsaw, Poland
}


\date{Received / Accepted }

\abstract{\oao\ is an accreting pulsar in an eclipsing 
binary system. We analyzed the INTEGRAL 
core program observations of this object and obtained 
the eclipse light curve in the soft gamma-ray band between 15 and 40 keV.
We note that the gamma rays from the pulsar allow to probe the density 
profile of the outer layers of the B supergiant companion.
We find that the  density profile  of the outer layer 
can be described by a power law with the index
 $\alpha = 8.5$. We also note that 
the fit hints toward smaller inclinations of the 
system within the allowed range $60^\circ<i<90^\circ$ \citep{1993ApJ...403L..33C}.
\keywords{pulsars: individual; gamma rays }
}

\maketitle

\section{Introduction}

The accreting pulsar \oao\ has been discovered with the Copernicus satellite by 
\citet{1978Natur.275..296P}. \citet{1979ApJ...233L.121W} found the $38.22$\,s 
pulsations  from the source. The pulsar spin  has been measured by a number 
of observatories and the spin history shows large spin-up and spin-down
episodes on a timescale between a few months and a year
\citep{1993ApJ...403L..33C,1997ApJS..113..367B}.
The companion has initially been identified as V861 Sco, yet this identification
has been proved to be incorrect \citep{1979ApJ...233L.121W,1980ApJ...236L.131A,1981ApJ...246..951B}.
The pulsar has been found to be a member of rare class of eclipsing objects
with an orbital period of 10.4\,d \citep{1993ApJ...403L..33C}, and the optical
companion identification was made possible only recently by the excellent X-Ray counterpart
location provided by Chandra
\citep{2002ApJ...573..789C}.
This identification lead to estimate of the distance at $6.4\pm 1.5$\,kpc, much
lower than the $11$\,kpc lower limit inferred previously with the assumption that the
neutron star is rotating at the equilibrium period.
The low energy X-ray spectrum of \oao\ is very absorbed, $N_H \sim 10^{23}$ cm$^{-2}$ 
\citep{1978Natur.275..296P,1990PASJ...42..785K},
because the source lies
at a low galactic latitude, and Ginga observations found  a soft excess below $3$\,keV
\citep{1990PASJ...42..785K,1979ApJ...233L.121W}, while \citet{1990PASJ...42..785K} found a
fluorescence line a $6.6$\,keV. 
The hard X-ray spectrum can be modeled by a power law steepening 
at energies above $20$keV, with a hint of a cyclotron line at $\approx 36$\,keV
\citep{1999A&A...349L...9O}.


\section{Data acquisition and processing}
 
INTEGRAL is a Gamma-ray satellite equipped with three 
high energy scientific instruments: SPI -- gamma spectrometer, IBIS -- gamma imager
and JEM-X -- X-ray monitor \citep{2003A&A...411L...1W}.
ISGRI is the low energy detector on board the IBIS/INTEGRAL telescope -- it is a position
sensitive $0.5$\,meter by $0.5$\, meter detector made of CdTe \citep{2003A&A...411L.131U}. 
It works in the energy range $15-1000$\,keV.
ISGRI consists of about 16\,384 single detector pixels.
The coded mask aperture is placed about 3 meters above the ISGRI detector.
The mask is made of the Tungsten segments: about 1 by 1 cm and 1.6\,cm thick,
in 95 by 95 segments system (half open), with the basic pattern of 53 by 53.
This coded aperture mask telescope collects gamma rays with
angular resolution better than 12 arcmin and with energy resolution
about 10\% at 100\,keV with about 2500\,cm$^2$ collecting area.

All the analyzed data come from ISGRI/IBIS detection layer.
We have selected the data intervals from the Galactic Plane Scan 
in the INTEGRAL core programme for which \oao\ was in the fully
coded field of view, i.e. less than $10^\circ$ from the axis.

\begin{figure}[t]
\includegraphics[width=0.9\columnwidth]{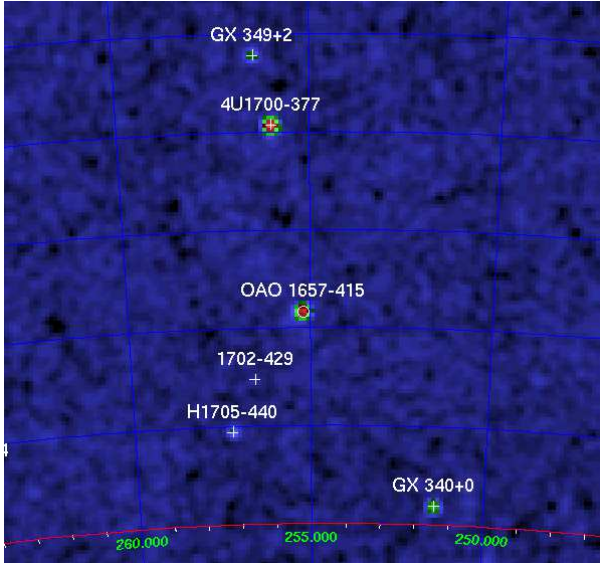}
\caption{The map of \oao\ region. The source is detected 
at the level of 122\,$\sigma$ for a total exposure of 24.5\,ks}
\label{map}
\end{figure}

This condition was satisfied by the observations 
from the the following INTEGRAL revolutions:
44-47, 49-56, 58-61, 63 (Spring 2003), and 
100, 101, 103, 116, 118, 119 (Autumn 2003)
The total  time of data acquisition was $\sim 900$\,ks.
The data has been processed using standard INTEGRAL software
provided by ISDC ({\tt http://isdc.unige.ch}) in version 3.0 with normal
configuration for the ISGRI pipeline.
A preliminary analysis of the data has been presented in \citet{2004sp-552.proc..453}.

Since we have used the data segments corresponding to 
different locations on the detector we had to model 
the off axis response of the IBIS/ISGRI detector. We have used 
the Crab observations during the INTEGRAL revolution 170
as our reference. We used two models of the detector angular response function: 
a wide area ($\pm$ 13$^\circ$ square around telescope axis) global fit
with a paraboloid summed with a two-dimensional Gaussian, and local flat surface triangulation.
Both gave similar results.

We have analyzed 480 science windows (INTEGRAL science window is 
a single exposure lasting of 20 minutes)
and the typical detection significance was 25
sigma. We have also detected the pulsations with the period of 37.16\,s.
We present the skymap of \oao\ region in Figure~\ref{map}.
The source is detected at the level of 122\,$\sigma$ in the 15$-$40\,keV 
energy band for a total exposure of 24.5\,ks \citep{2004sp-552.proc..453}.

\section{Eclipse modeling}

\begin{figure}[t]
\includegraphics[angle=-90,width=0.9\columnwidth]{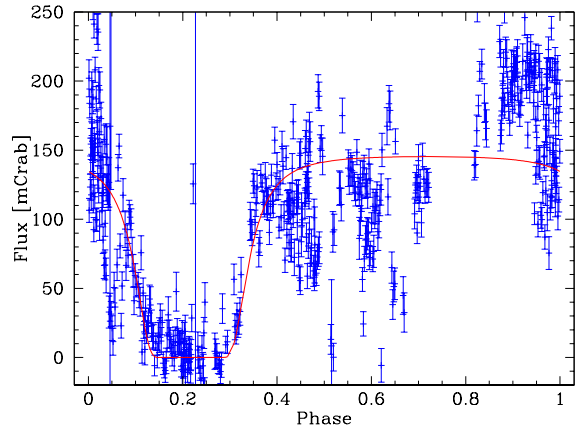}
\caption{OAO lightcurve with the best fit of the eclipse model (see text for details)}
\label{lc}
\end{figure}

The brightness of the source varies from 200-300\,mCrab to 
nearly  0 between different observations. 
We have folded the observations with the orbital period
of the pulsar of $10.44$\,days discovered by \citet{1993ApJ...403L..33C}.
This revealed a beautiful eclipse profile, as can be seen 
in Figure~\ref{lc}. Epoch of the eclipse centre is JD 2\,452\,705.9.

The eclipse in the lightcurve is clearly visible. 
What is interesting is that the eclipse ingress and regress are gradual,
and last almost a day, with the totality, ingress and regress,
of almost two days.
A pulsar is essentially a point source in comparison with 
the B supergiant companion with a radius of $\approx 30\,R_\odot$.
Such a gradual eclipse is expected when 
the sizes of the sources are comparable, or if the 
eclipsing object is diffuse and has a low optical thickness to the 
gamma rays of the eclipsed source.
In the case of \oao\ light curve we have to consider 
this latter case.

To model the eclipse we have assumed that the 
density profile of the B supergiant companion can be described
as a power law:
\begin{equation}
\rho(r)=\rho_0   \left( {r\over 30\,R_{\sun} } \right)^{-\alpha}.
\label{density}
\end{equation}
For a given orbital phase $\phi$ and assumed inclination  $i$ we 
can calculate the  the column density through the eclipsing star
by integrating the density profile:
\begin{equation}
n_e(\phi) = \int_{r_\parallel (\phi)}^\infty  dr' \rho(\sqrt{r_{\perp}(\phi)^2
+r'^2})
\end{equation}
where $r_\parallel(\phi)$ is the projection of 
distance between the pulsar and the companion along the line of sight,
and  $r_\perp(\phi)$ is the projection of this distance on the plane perpendicular 
to the line of sight.
At the energies $15-40$\,keV the main opacity through the star 
is due to electron scattering.
We treat the  radiative transfer approximately and we model the eclipsed 
flux from  the pulsar as
\begin{equation}
F(\phi) = F_0 \exp(-n_e(\phi-\phi_0)\sigma_T)
\end{equation}
where $F_0$ is the flux from the pulsar,
$\phi_0$ is the center of the eclipse
and $\sigma_T$ is the Thompson cross section.
Altough the source is intrinsically variable (20-40\%), 
we put $F_0$ constant because the modeled eclipse light curve is a superposition of many
data sets covering randomly the whole range of the orbital phase.
Such an approach should not have an important impact on the
fitted values of the eclipse parameters ($i$, $\phi_0$, $\rho_0$, and $\alpha$), but
can increase the incertitude on $F_0$.
In the model we use the orbital elements inferred 
earlier from the BATSE data \citep{1993ApJ...403L..33C}.
The model has five free parameters: 
$i$, $\phi_0$, $F_0$, $\rho_0$, and $\alpha$.
  
The data analysis yielded the statistical errors. We estimate 
the additional systematic errors by requiring that the data in the
total eclipse part (phases from 0.13 to 0.30
) 
are consistent with zero flux.
To this goal we add in quadrature a systematic error 
$\sigma_{sys} =15.5$\,mCrab. 
This is a minimal value which leads $\chi^2/DOF=1$ in this phase interval 
where the flux is zero during the total eclipse.

To find the best fit model we use the $\chi^2$ minimalization. 
We have scanned the parameter space, with the  restriction
$60^\circ< i <90^\circ$, as inferred from the timing analysis 
by \citet{1993ApJ...403L..33C}. 
A best fit model light curve is presented 
as the solid line in Figure~\ref{lc}. 
The model curve  correspond to the model with the following parameters:
$i=60^\circ$, $F_0=147$\,mCrab, $\alpha=8.6$, $\phi_0=0.218$, and
$\log \rho_0=12.34$, where $\rho_0$ is in units of cm$^{-3}$.

\begin{figure}[t]
\includegraphics[width=0.9\columnwidth]{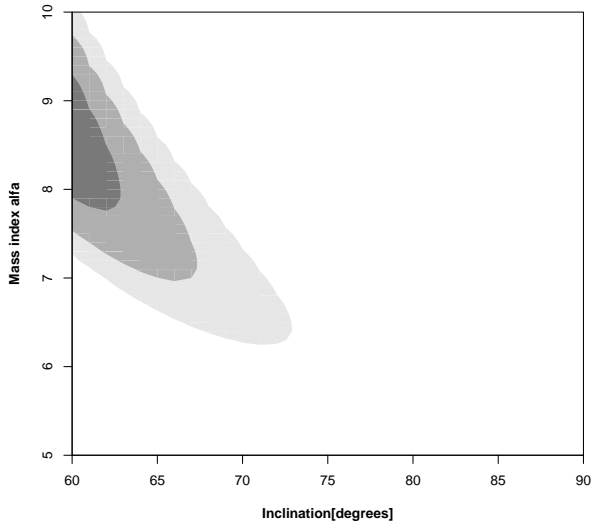}
\caption{The gray contours enclose the regions containing
$68$, $95$, and $99.9$ percent probability. The 
index $\alpha$ increases with decreasing inclination angle}
\label{ta}
\end{figure}

\begin{figure}[t]
\includegraphics[width=0.9\columnwidth]{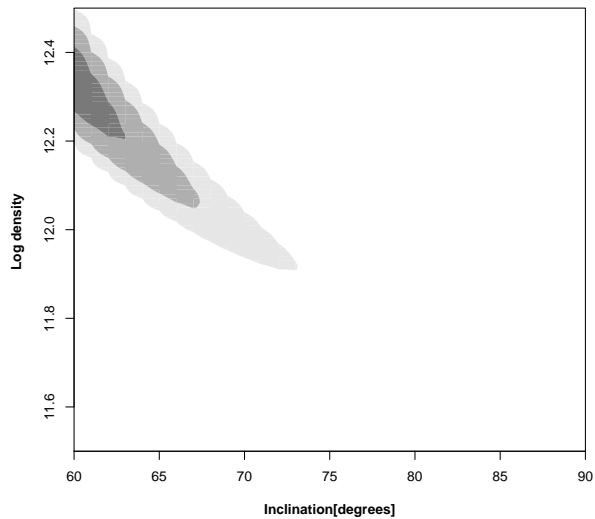}
\caption{The gray contours enclose the regions containing
$68$, $95$, and $99.9$ percent probability. The 
density $\rho_0$ increases with decreasing inclination angle}
\label{tr}
\end{figure}

In order to find the confidence regions on the parameters
we assume that the probability density in the parameter
space is proportional to $\propto \exp(-\chi^2(p_i))$ where 
$p_i$ are the parameters. 
Such probability density is normalized.
We marginalize over the remaining parameters to find the 
probability distributions for each parameter, or two dimensional maps
of probability density. 
The best fit values with the $1\sigma$ confidence contours 
obtained using the above procedure are 
$\alpha=8.5\pm^{1.2}_{1.5}$, $\log \rho_0=12.31\pm^{0.15}_{0.30}$,
the center of the eclipse is $\phi_0=0.218\pm 0.001$. The formal 
best fit value for the flux $F_0$ is $F_0=147\pm 1$\,mCrab, and for the 
inclination $i=60^\circ\pm_{0}^4$. The best fit values 
of the parameters corresponds to the peak of the probability 
distribution and the $1\sigma$ describe the region
containing 68\% probability around this peak. The formal 
error on inferred value of the inclination angle reflects the fact that we
considered only the angles above $60^\circ$ and the
best fit corresponded to the lowest angle considered.

The relatively large uncertainties in determination of
the mass index $\alpha$ and the fiducial density $\rho_0$ 
are due to the fact that these quantities are correlated. 
In Figures \ref{ta} and \ref{tr} present  the regions containing $68$, $95$, 
and $99.9$ percent of the probability in the space spanned by 
the index $\alpha$ and the inclination angle (Figure~\ref{ta}), 
and by the density $\rho_0$ and the inclination angle (Figure~\ref{tr}). 
Both the mass index $\alpha$ and the density $\rho_0$ are 
anticorrelated with the inclination angle of the binary. 
The increase of the density parameter is due to the fact that 
as we decrease the inclination angle the eclipse probes farther 
radii from the stellar center, while $\rho_0$ is defined as density 
at a constant radius of $30R_\odot$.
The index $\alpha$ increases since at higher inclinations
the neutron star is eclipsed along a chord line and the density
difference leading to the observed eclipse corresponds to smaller
difference of the radii from the stellar center. 
Therefore also the index $\alpha$ is correlated with 
the density parameter $\rho_0$.

\section{Summary and discussion}

The INTEGRAL observations of the eclipse
are the first result where a stars outer layers can be probed 
directly by the external gamma ray source. 
The mass enclosed in the region that we probe 
can be estimated by integrating equation \ref{density}
starting from the point where the eclipse becomes total.
The result depends on the assumed inclination and varies 
from $5\times 10^{-7}\,M_\odot$ for $i=90^\circ$ to
$1.6\times 10^{-5} \,M_\odot$ for $i=60^\circ$.
The thickness of the layer that is probed with this method
also depends on the assumed inclination. 
Assuming that the width of the total eclipse is $0.14$ 
we estimate that with the inclination of $90^\circ$ we probe 
down to the radius of $\approx 20\,R_\odot$ from the center, 
while at the assumed inclination $i=60^\circ$ we see 
the structure above $28\,R_\odot$.

The outer layer of the star is the location of the 
acceleration of the wind. 
Our results can be used along with the continuity equation 
to estimate the wind velocity profile: $v(r)\propto r^{\alpha-2}$. 
This estimate uses an assumption that we already see the accelerating
layer. In reality the wind may be taking off in the upper
portion of the atmosphere which is already optically thin 
to gamma rays. 
 
Finally we note that the flux outside the eclipse seems
to increase by a factor of $\approx 2$ before entering the eclipse,
i.e. between the phase 0.4-0.6 and 0.8-1.0. 
While the data looks particularly noisy it seems interesting to investigate 
the possibility that the variation of the flux is due to the structure of the 
outgoing wind. 
The accreting neutron star may leave a hollow
trace in the wind and such a trace could correspond 
to decrease of density during a particular phase, when the pulsar
is viewed through hole in the wind material.

Summarizing, we have presented 
the first well observed case of a gamma-ray eclipse.
We show that the eclipse lightcurve allows to study
the structure of the outer layers of the companion.
INTEGRAL is particularly well prepared for such studies
as it allows long duration monitoring of sources.

\begin{acknowledgements}
INTEGRAL is an ESA project with instruments and science data center
funded by the ESA members (especially the PI countries: Denmark, 
France, Germany, Italy, Switzerland, Spain), Czech Republic, and Poland, 
and with the participation of Russia and USA.
This research was supported by the KBN grants 2\,P03D\,001\,25 and PBZ-KBN-054/P03/2001.
\end{acknowledgements} 
 

\end{document}